\documentclass[a4paper,fleqn,usenatbib,useAMS]{mnras}

\usepackage{graphicx}	
\usepackage{amsmath}	
\usepackage{amssymb}	
\usepackage{multicol}        
\usepackage{bm}		
\usepackage{pdflscape}	
\usepackage{journals}

\def\M{M\,}
\def\N{NGC\,}
\def\NGC{NGC\,}
\def\HI{HI}
\def\Ms{$\textrm{M}_{\odot}$}
\def\kms{$\textrm{km~s$^{-1}$}$}
\def\nb{\textsc{nbursts}}

\usepackage[T1]{fontenc}
\usepackage{ae,aecompl}

\title[The Diversity of Thick Discs]{The Diversity of Thick Galactic Discs}

\author[A.Kasparova et al.]{Anastasia V. Kasparova,$^{1}$\thanks{Contact e-mail: \href{mailto:anastasya.kasparova@gmail.com}{anastasya.kasparova@gmail.com}},
Ivan Yu. Katkov,$^{1}$ 
Igor V. Chilingarian,$^{2,1}$ 
\newauthor
Olga K. Silchenko,$^{1,3}$ 
Alexey V. Moiseev$^{4,1}$ 
and Svyatoslav B. Borisov$^{1,5}$.
\\
$^{1}$Sternberg Astronomical Institute, Moscow M.V. Lomonosov State University, Universitetskij pr., 13,  Moscow, 119992, Russia\\
$^{2}$Smithsonian Astrophysical Observatory, Harvard-Smithsonian Center for Astrophysics, 60 Garden St. MS09, Cambridge, MA 02138 USA\\
$^{3}$Isaac Newton Institute of Chile, Moscow Branch,  Universitetskij pr., 13,  Moscow, 119992, Russia\\
$^{4}$Special Astrophysical Observatory, Russian Academy of Sciences, Nizhnij Arkhyz, 369167, Russia\\
$^{5}$Department of Physics, Moscow M.V. Lomonosov State University, 1, Leninskie Gory, Moscow, Russia, 119991\\
}

\pubyear{2016}

\begin{document}
\label{firstpage}
\pagerange{\pageref{firstpage}--\pageref{lastpage}}
\maketitle

\begin{abstract}
Although thick stellar discs are detected in nearly all edge-on disc
galaxies, their formation scenarios still remain a matter of debate.  Due to
observational difficulties, there is a lack of information about their
stellar populations.  Using the Russian 6-m telescope BTA we collected deep
spectra of thick discs in three edge-on S0-a disc galaxies located in
different environments: \N4111 in a dense group, \N4710 in the Virgo
cluster, and \N5422 in a sparse group.  We see intermediate age ($4-5$~Gyr)
metal rich ([Fe/H]\ $\sim -0.2\dots0.0$~dex) stellar populations in 
\N4111 and \N4710.  On the other hand, \N5422 does not harbour
young stars, its disc is thick and old (10~Gyr), without evidence for a second component, and its
$\alpha$-element abundance suggests a $1.5-2$~Gyr long formation epoch
implying its formation at high redshift.  Our results suggest the diversity of
thick disc formation scenarios.
\end{abstract}




\section{Introduction}

Many important aspects of galaxy evolution remain not fully understood
despite the great progress in the astronomical instrumentation and
increasing resolution and complexity of numerical simulations.  One such
unsolved problem is the formation of thick discs, important and widespread
structural elements of spiral and lenticular galaxies
\citep{Burstein79}. 
Manifested by the exponential excess of
light in edge-on disc galaxies at large distances above the main disc
plane, they are found in most if not all cases
\citep{DalcantonBernstein2002}.  The Milky Way contains a thick
stellar disc component with the scale-height of $\sim1$~kpc that harbours
older and more metal poor stars compared to the starforming thin disc. 
Lenticular galaxies sometimes possess only old thick discs \citep{McDermid+15}
being consistent with a scenario where they formed thick
discs at high redshifts \citep[see details in][]{Silchenko2012} 
and then have never acquired thin discs and become \textit{normal} spirals 
in contrast to the formation scenario of the lenticular galaxies through
quenching of spirals \citep{Larson+80}.

Several thick disc formation scenarios have been proposed. (i)~Thick discs
formed rapidly at high redshifts as a result of high density and velocity dispersion of
gas in the early Universe \citep{Elmegreen2006, Bournaud2009},
and thin discs formed consequently by gas accretion from filaments (see
\citealp{CMG97,Combes14} and references therein) or minor \textit{wet} mergers
\citep{Robertson+06,Silchenko2011NGC7217}. This scenario should produce old
$\alpha$-enhanced thick discs without notable metallicity gradients.  
(ii)~Thick discs can be formed via secular thin disc flaring as a result of
radial migration of stars \citep{Schonrich2009,Loebman2011,RoskarDebattista+2013}. 
Some scenarios predict specific radial and vertical stellar population
patterns, for example, a negative radial age gradient above the principal disc plane
\citep{Minchev2015}. 
(iii)~Primordially thin discs can get dynamically heated by satellite flybys
\citep{Quinn93,Qu+2011a} and/or minor mergers.  

Detailed studies of internal
kinematics and stellar populations in thin and thick discs will help us
to choose the scenario.
In external galaxies
which cannot be resolved into stars, integrated light spectroscopy remains
the only observational technique for thick disc studies.  However, it
represents a significant challenge because of typical low surface
brightnesses of thick discs.

We carried out deep long-slit spectroscopic observations for a sample of edge-on 
disc galaxies and derived spatially resolved stellar kinematics and star
formation histories. In this \textit{letter} we present the first results on
three S0-a galaxies which, as we show, prove the diversity of the thick disc
formation scenarios.

\section{The Sample and the Data}

\subsection{The Sample}

We chose three edge-on galaxies in different environments: \N4111, \N4710
and \N5422 morphologically classified as \textit{S0-a} by
Hyperleda \citep{Makarov2014}. 

\begin{itemize}
\item \textbf{\N4111} ($M_J = -22.40$~mag) is a member of the Ursa Major galaxy
group that is known to contain a common extended {\HI} envelope
\citep{Wolfinger2013}.  The distances from {\N4111} to the nearest
neighbours are about $30-40$~kpc \citep{Pak2014,Karachentsev2013}.  We adopt
the distance 15~Mpc that corresponds to the spatial scale
72.7~pc~arcsec$^{-1}$ \citep{Tonry2001}.
\item \textbf{\NGC4710} ($M_J = -22.56$~mag) is located in the Virgo cluster
outskirts ($d=16.5$~Mpc by \citet{Mei2007}, spatial scale
80.0~pc~arcsec$^{-1}$).  Its projected distance to~\M87 is about 5.4~deg
or 1.6~Mpc \citep{Koopmann2001}.  A dusty disc is observed in the central 2~kpc region
dominated by an X-shaped structure, that is traditionally explained as an
edge-on bar \citep{BF99}. 
\item \textbf{\NGC5422} ($M_J = -22.81$~mag) is a member of the sparse
\N5485 galaxy group dominated by lenticular galaxies.  It is the most
luminous and the most distant object in our sample ($d=30.9$~Mpc by \citet{Theureau2007}, spatial scale
150~pc~arcsec$^{-1}$). It
possesses a large gaseous disc tilted by some $\sim 5$~deg with respect
to the stellar disc.
\end{itemize}

\subsection{New Observations and Archival Data}

We obtained deep spectroscopic observations of all three galaxies with the
universal spectrographs SCORPIO \citep{Afanasiev2005} and
SCORPIO-2 
\citep{AfanasievMoiseev2011} at the Russian 6-m BTA telescope using the
1~arcsec wide 6~arcmin long slit.  For every galaxy we observed: (i)~a~major
axis in order to get the information on a thin disc for \N4111 and \N4710,
and the mid-plane of the thick disc in \N5422 and (ii)~a~region parallel to
the major axis offset by $5-7$~arcsec that corresponds to $0.36-1$~kpc above
the mid-plane in order to probe thick discs.  We observed (see
Table~\ref{tablog}) the mid-plane of
\N4111 with the SCORPIO in the wavelength range $4800 - 5600$~\AA\ with the spectral
resolution $\sim 2.2$~\AA\ $=55$~\kms.  The remaining 5~datasets were obtained  using SCORPIO-2 at slightly lower spectral resolution ($\sim 3.8$ \AA\
$=95$~\kms~at 5100 \AA) at the broader wavelength range ($3600 - 7070$~\AA).

\begin{table}
\caption{Long-slit spectroscopy of the sample galaxies.\label{tablog}}
\scriptsize
\begin{tabular}{lllcccc}
    \hline     
NGC & Date & z-offset &  $PA$ & Sp. range & $T_{exp}$ & Seeing \\
 & 	   & arcsec/pc  & deg & \AA & sec & arcsec \\
\hline   
4111 & 21/05/09  & 0/0  & $150$ & 4825-5500 & 8400 & 1.3\\
4111 & 24/04/15  & 5/364 & $150$ & 3600-7070 & 5600 & 1.0\\
4710 & 24/04/15  & 0/0  & $27.5$ & 3600-7070 & 3600 & 1.2\\
4710 & 24/04/15  & 7/560 & $27.5$ & 3600-7070 & 7200 & 1.1\\
5422 & 24/04/12 & 0/0  & $151.4$ & 3600-7070 & 3600 & 2.5\\
5422 & 25/04/15 & 7/1049 & $151.4$ & 3600-7070 & 8400 & 1.5\\
\hline
\end{tabular}
\end{table}

We reduced our spectroscopic observations with our own {\sc idl}-based
reduction pipeline.  We estimated the night sky background from outer
slit regions not covered by our target galaxies and an optimized sky
subtraction technique that takes into account spectral resolution variations
along the slit
\citep{skysubtr_adass_proc2011,katkov2014_ilgpop_skysubtr}.

All three galaxies were observed with the Infrared Array Camera (IRAC) at
Spitzer Space Telescope in the imaging mode at 
wavelengths 3.6~$\mu$m and 4.5~$\mu$m.  
We fetched fully reduced 3.6~$\mu$m images from
the Spitzer Heritage Archive\footnote{\url{http://sha.ipac.caltech.edu/applications/Spitzer/SHA/}}.

\subsection{Data Analysis}

\citet{vdKS81} have shown that in case an isothermal disc in the equilibrium state the vertical disc density profiles are described
by the law $I = I_0\ \textrm{sech}^2(z/z_0)$, where $I_0$ is the central
intensity and $z_0$ is the disc scale-height.  
We fitted vertical profiles obtained by averaging Spitzer IRAC
images along the radius using models including one and two components.
We set central positions of both components to be the same in the case of two-component
fitting but left the position itself a free parameter.  Our results for
\N4111 and \N5422 quantitatively agree with those presented in the
S4G survey \citep{Salo2015}, however the results for \N4710 decomposition were not presented there.

\begin{figure*}
\centering
\includegraphics[width=0.85\textwidth]{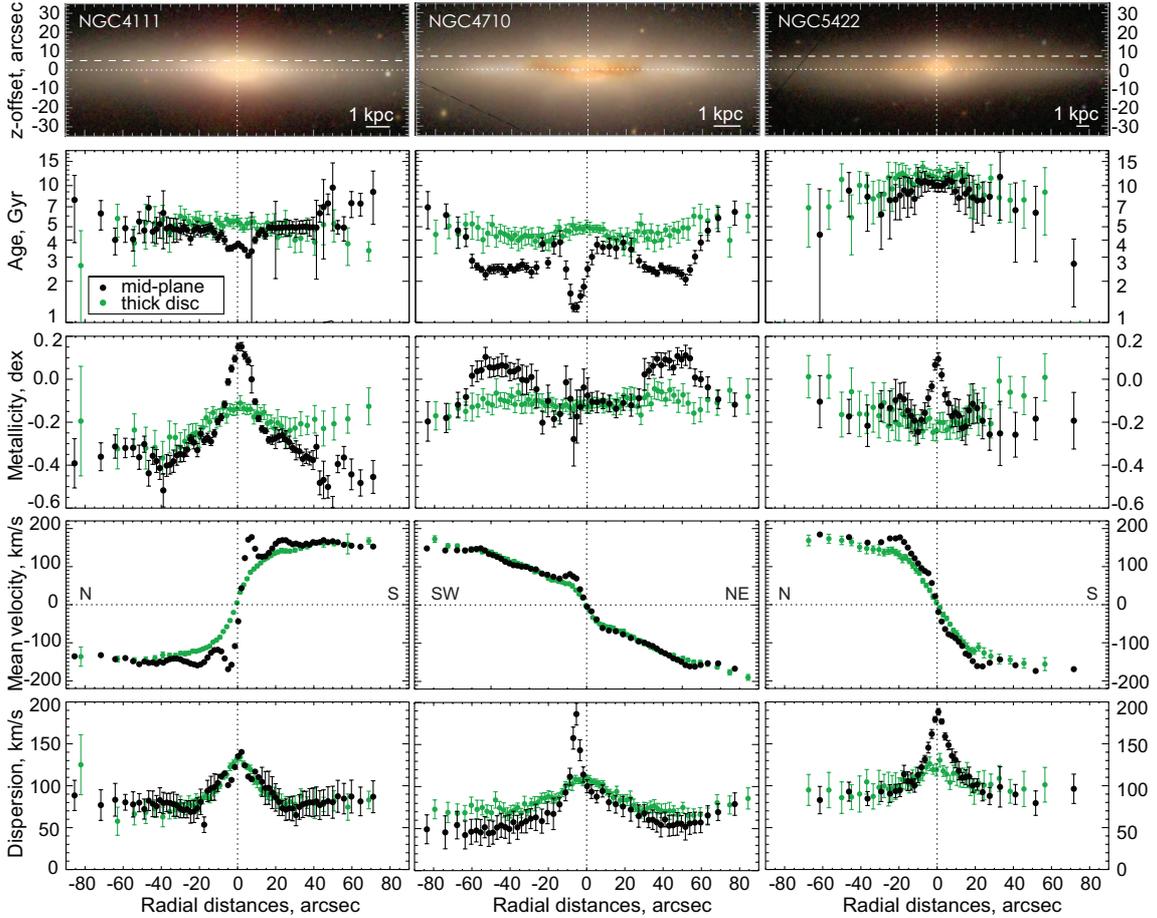}
\caption{Stellar populations and internal kinematics of three edge-on disc
galaxies. Panels (top to bottom): the slit positions overplotted on SDSS color
images, SSP equivalent ages, metallicities, radial velocities and line-of-sight velocity dispersions. Black
and green symbols denote mid-planes and thick discs. 
}  
\label{fig1}
\end{figure*}

To derive internal kinematics and stellar population properties (mean ages
and metallicities [Fe/H]) of thick and thin discs we first binned our long slit
spectra in the spatial direction with the adaptive binning algorithm in
order to reach the minimal signal-to-noise ratio $S/N=30$ per bin per
spectral pixel in the middle of the wavelength range.  Then, in every bin we
applied the \nb\ full spectral fitting technique \citep{nbursts_a,nbursts_b}
with a grid of high resolution stellar {\sc pegase.hr} \citep{pegasehr}
simple stellar population (SSP) models based on the ELODIE3.1 empirical
stellar library \citep{elodie3.1}.  The \nb\ technique implements a pixel
space $\chi^2$ minimization algorithm where observed spectrum is
approximated by a stellar population model broadened with parametric
line-of-sight velocity distribution (LOSVD) and multiplied by polynomial
continuum (10th degree in our case) to take into account dust attenuation
and/or possible flux calibration imperfections in both observations and
models.  We used Gaussian parametrization of the stellar LOSVD, except our
higher resolution mid-plane spectra of \N4111 where we exploited the
standard Gaussian-Hermite parametrization \citep{gausshermite}.

The [Mg/Fe] abundance ratio which allows one to estimate the duration of the star
formation epoch is fixed to the solar value in the {\sc pegase.hr} model
grid. Therefore, we measured the Mg$b$, Fe$5270/5335$ Lick indices
\citep{Worthey1994} and derived [Mg/Fe] ratios in several radial bins along
the slit using $\alpha$-variable models from \citet{Thomasstpop}.

\section{Results and Discussion}

\begin{figure}
\centering
\includegraphics[scale=0.56]{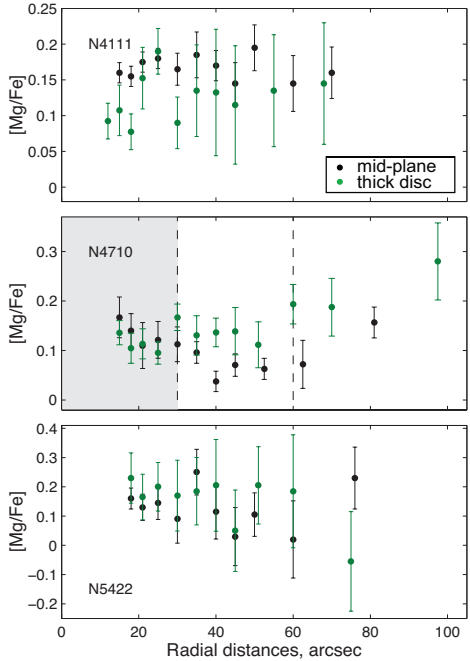}
\caption{The [Mg/Fe] values averaged in radial bins. Black and green dots are for thin and thick discs, respectively. 
Dotted lines denote the three regions of \N4710, shaded gray is a bar dominated area, then the
\textit{inner} disc at $30<r<60$~arcsec, and the thick disc dominated area. 
\label{Mg2Fe}}
\end{figure}

\subsection{\N4111}

In Fig.~\ref{fig1} (left plots) we show the SSP-modelling results of
\N4111 stellar population and kinematic data. \citet{Comeron2014} have estimated
the thin and thick disc masses to be $15\pm3\cdot10^9$~\Ms\ and $5\pm1\cdot10^9$~\Ms\ respectively, and \citet{Comeron2012}
have demonstrated that the thick disc of \N4111 dominates the luminosity at
$z>25.4$~arcsec ($\sim1.8$~kpc) in the inner region. However, according to
their models, already at $r=40$~arcsec the thick disc contribution
represents about 80~per~cent of the total luminosity fraction in the
mid-plane and reaches 90~per~cent at $z=5$~arcsec were we placed the~slit. 
Our data extend out to 6.2~kpc from the centre that corresponds to 10.0
and 2.7 thin and thick disc scale-lengths respectively.
Nevertheless, we do not see significant differences of stellar populations
in the outer regions of the \N4111 discs. The [Mg/Fe] values for both disc
components are about +0.15~dex and consistent within 0.03~dex (Fig.~\ref{Mg2Fe}, top panel).  Stellar
ages of the two layers are almost identical $\sim5$~Gyr. Moreover, the
stellar velocity dispersion profiles are similar too.  The slight asymmetry of
the thick disc stellar population profiles in the outer regions is not
statistically significant.  With the exception of the central region where
our mid-plane data probes the bulge and a kinematically decoupled
co-rotating inner disc (within 10~arcsec), we see a completely flat age
profile.  Metallicity has a slight negative gradient and decreases from
[Fe/H]\ $=-0.2$~dex at $r=15$~arcsec to $-0.4$~dex in the outer regions
of both discs.

\subsection{\N4710}

Our data extends to 85~arcsec (6.8~kpc) and we can identify at least three
distinct regions in \N4710 by radius: (i)~an~inner dusty region
within 30~arcsec that corresponds to the outer edge of a~clearly
seen X-shaped structure, (ii)~a~bright thin blue disc extending to 60~arcsec, and (iii)~weaker outer regions.  Mid-plane age and metallicity
profiles confirm the stepped structure (black points on the centre panels of
Fig.~\ref{fig1}).  A slight asymmetry within the central region corresponds
to the bright spot distinctly seen on optical images of \N4710 south-west of
its centre.  It is probably a giant star formation region that shines
through unevenly distributed dust.  Outside the dusty region, between 30 and
60~arsec, we see a young ($\sim2.5$~Gyr) metal-rich ($\sim+0.1$~dex)
component without significant radial stellar population gradients.  Outside
60~arcsec the stellar age abruptly becomes significantly older and reaches
$6$~Gyr while the metallicity decreases to $\sim-0.17$~dex (e.g. almost
half of that inside 60~arcsec).  At the same time, the thick disc of \N4710
(at 560~pc above the mid-plane) possesses significant gradients neither in
age nor in metallicity, which coincide with the values in
both outer ($>60$~arcsec) and inner ($<30$~arcsec) regions in the mid-plane,
$t\approx4\dots5$~Gyr, [Fe/H]\ $\approx -0.15$~dex.

\begin{figure}
\centering
\includegraphics[scale=0.57]{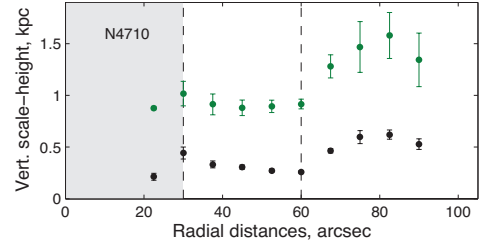}
\caption{Scale-height radial variations of the \N4710 thin and thick discs. The designations are as in Fig.~\ref{Mg2Fe}.
\label{fig2}}
\end{figure}

The vertical profile decomposition reveals two components with the
scale-height values related as $\sim 1:3$ 
at all radii outside the bar dominated region.  At $r>60$~arcsec both components are flaring and become twice as thick while their ratio still holds (Fig.~\ref{fig2}).

\N4710 is the only object in our small sample where we detect statistically
significant differences in [Mg/Fe] abundance ratios between the thin and thick
disc components and where we see its radial gradients (see Fig.~\ref{Mg2Fe},
middle panel).  While the values coincide at slightly supersolar
[Mg/Fe]\ $\approx+0.15$~dex inside 30~arcsec, the region dominated by the bar,
the thin disc $\alpha$-enhancement drops to $+0.05$~dex at $30<r<60$~arcsec
and then raises again at $r>70$~arcsec to $+0.15$~dex, while thick disc stays
nearly constant ($+0.15\dots0.20$~dex) at all radii.

Also, \N4710 is the only galaxy out of three where dust can affect our 
stellar population analysis. This is true for the inner region ($r<30$~arcsec) 
where we see the uneven distribution of dust in the archival optical (HST) and far infrared (Hershel) images.
However, we do not consider this area in our thick disc analysis.
Beyond 30~arcsec, we can neglect the dust effects because it is not detected
in any substantial amount on Herschel images.

\subsection{\N5422}

\N5422 has a single disc component as evident from its vertical density profile
analysis.  The two-dimensional decomposition by the S4G project reveals a
bulge and one \textit{thick} disc with the scale-length and scale-heights
24.3 and 6.7~arcsec  
\citep{Salo2015}.  The disc warp
semi-amplitude reaches 110~pc at $r=3$~kpc.  \N5422 also possesses a
large scale non-starforming gaseous disc slightly inclined at 5~deg to the
main plane manifested by weak emission lines.

For comparison with the mid-plane of \N5422 we placed a slit at 7 arsec
($\sim1050$~pc) above it.  Our profiles of kinematics and stellar population
parameters extend to 70~arcsec (10.5~kpc). 
We expect that outside $r=20$~arsec the bulge influence becomes negligible.  Not
surprisingly, in those regions our analysis demonstrate stellar population
properties and stellar velocity dispersion radial profiles consistent within
uncertainties (Fig.~\ref{fig1}, right plots).  \N5422 is the only galaxy
were we see an old thick disc $\sim8\dots11$~Gyr with slightly subsolar
$\sim-0.2$~dex metallicity and a moderate $\alpha$-enhancement
([Fe/H]\ $\sim0.15$).  The age and metallicity gradients in \N5422 are
not statistically significant. 

\subsection{The Diverse Origin of Thick Discs}

Several processes may act simultaneously during the formation of
structures that we observe how as thick discs.  When we extend our sample,
our high quality observational data will allow us to evaluate the
significance of every particular thick disc formation scenario not only for
specific objects but also for different galaxy types.  Presently, all three
objects we describe are classified as S0-a galaxies, which certainly cannot
give a full picture of the thick disc formation in late type spiral
galaxies.  Nevertheless, the assumption that S0's may be disc systems depleted their gas by star formation \citep[see e.g.][]{Larson+80} implies a possible evolutionary link between thick discs
across the Hubble sequence.

According to theoretical predictions, we do not expect that in lenticular galaxies without strong spiral arms, the radial migration should play an
important role. Given our very limited sample, it is difficult to 
test the relevance of the radial migration scenario from observations.
One of the reasons is that depending on the initial distribution of the radial
metallicity, very different radial profiles of stellar population parameters
can emerge \citep[see e.g.][]{Curir2012}.  
Nevertheless, the models by \citet{Minchev2015} predict a notable negative
age gradient in thick discs because younger stars migrate further from the disc plane at
larger radii.  Our data extend substantially further in terms of surface
brightness and, consequently, radial distances than in previous studies 
\citep{YoachimDalcanton2008,Comeron+2015}, therefore the flat radial age
profiles which we observe can rule out this formation model. On the other
hand, this situation may occur if the radial migration was strong enough to
flatten radial stellar population gradients in both thin and thick discs.

The rapid turbulent thick disc formation by \citet{Bournaud2009}
looks plausible only for \N5422. This galaxy has a moderately
$\alpha$-enhanced very old stellar population in its disc
that correspond to the duration of the star formation epoch of $1.5-2$~Gyr
\citep{Thomas+05}, if it had started to form at $z\approx3$ and was finished by
$z\approx1.5$.

At the same time, we cannot clearly assess the importance of
other thick disc formation mechanisms, such as minor mergers or
accretion events, for all three galaxies.  For example, very moderate
negative or flat metallicity gradients in our thick and thin discs support
the scenario of the disc thickening by minor mergers, because such events
are expected to flatten metallicity gradients as demonstrated by numerical
simulations \citep{Qu+2011a} available in the {\sc galmer} database
\citep{Chilingarian2010}.

In addition to the above, we emphasize some environment related
aspects, which may explain the formation and evolution features of disc
subsystems under consideration.

The apparent multi-component structure in the mid-plane of \N4710 suggests a
complex evolution history.  We have to keep in mind that this galaxy belongs
to the Virgo cluster.  Despite of the fact that it is nowadays located at a
significant distance from \M87 ($\sim1.6$~Mpc), we have a reason to
believe that \N4710 already passed the cluster center in the past. 
According to \citet{Koopmann2001}, there is an abrupt truncation of
GALEX UV and H$_{\alpha}$ fluxes, tracers of recent and ongoing star formation,
beyond $r=60$~arcsec.  Moreover, the map of atomic hydrogen demonstrates the
truncated profile and H{\sc i} (as well as dust) is detected only within the
inner 60~arcsec \citep{Serra2012}.  
We believe that we should see a similar
end-product, if an initially thick disc had been forming and growing until low
redshift ($z=0.5$ in this case) similarly to \N4111, and then the thin disc
growth was fueled by additional gas in
much larger amounts than in \N4111, which caused intense star formation and
self-enrichment. 
Later, the ram pressure stripping \citep{GunnGott1972} by
the hot intracluster gas near the Virgo cluster center strips the outer part
($r>60$~arcsec) of the thin gaseous disc and quenches the star formation. 
The disc flaring would quickly occur beyond this radius because there will
be no more gas and dynamically cold stars forming in the mid-plane~---
exactly as we observe in \N4710.

\N5422 does not have a young stellar component at all. 
It is a member of the sparse \N5485 group where at least 4 out of 5 other luminous galaxies
are lenticulars with discs older than $t>10$~Gyr \citep{Peletier1996,Peletier1999}.  They
all might have formed at $z=2\dots3$ as \textit{normal} discs and then exhausted
their gas reservoirs around $z=1$ and stopped forming stars.  Since then,
the group must have been evolving only by dry mergers.  The \N5422 disc
looks very similar to the face-on lenticular galaxy \N6340
\citep{chil_6340_skysubtr} that also lives in a sparse group.

After all, the three examples of edge-on thick discs revealed the diversity
of properties that suggest a broad spectrum of their formation scenarios.
Further clarifications on preferred scenarios will be obtained when our
sample of deep spectroscopic observations extends.

\section*{Acknowledgments} 
We are grateful to Anatoly Zasov for productive discussions and we thank the anonymous referee for useful comments. 
The Russian 6m telescope is exploited under the financial support by the
Russian Federation Ministry of Education and Science (agreement
No14.619.21.0004, project ID RFMEFI61914X0004).
Our deep spectroscopic observations of thick discs are
supported by the Russian Science Foundation project 14-22-00041.  
The archival data analysis is supported by
the grants MD-7355.2015.2 and RFBR 15-32-21062.  The interpretation of
the spectroscopic data was performed at the annual Chamonix workshop
supported by the joint RFBR-CNRS project 15-52-15050. The authors acknowledge partial support from the M. V. Lomonosov Moscow State University Program of Development.



\bibliographystyle{mnras}
\bibliography{Thicks_disk_art}  


\bsp	
\label{lastpage}
\end{document}